\documentclass[fleqn,10pt]{wlscirep}

\usepackage{braket}
\usepackage{amsmath}
\usepackage{float}

\title{Triplet superconductivity in coupled odd-gon rings}

\author[1,2,*]{Sahinur Reja}
\author[3,4,+]{Satoshi Nishimoto}

\affil[1]{Department of Physics, Indiana University, Bloomington, Indiana 47405, USA}
\affil[2]{School of Mathematics and Physics, The University of Queensland, Brisbane, Queensland 4072, Australia}
\affil[3]{Technical University of Dresden, Department of Physics, Dresden, 01069, Germany}
\affil[4]{IFW Dresden, Institute for Theoretical Solid State Physics, Dresden, 01069, Germany}

\affil[*]{s.reja@uq.edu.au}
\affil[+]{s.nishimoto@ifw-dresden.de}

\begin{abstract}
Shedding light on the nature of spin-triplet superconductivity has been a long-standing quest in condensed matter physics since the discovery of superfluidity in liquid $^3$He. Nevertheless, the mechanism of spin-triplet pairing is much less understood than that of spin-singlet pairing explained by the Bardeen-Cooper-Schrieffer theory or even observed in high-temperature superconductors. Here we propose a versatile mechanism for spin-triplet superconductivity which emerges through a melting of macroscopic spin polarization stabilized in weakly coupled odd-gon (e.g., triangle, pentagon, etc) systems. We demonstrate the feasibility of sustaining spin-triplet superconductivity with this mechanism by considering a new class of quasi-one-dimensional superconductors A$_2$Cr$_3$As$_3$ (A=K, Rb, and Cs). Furthermore, we suggest a simple effective model to easily illustrate the adaptability of the mechanism to general systems consisting of odd-gon units. This mechanism provides a rare example of superconductivity from on-site Coulomb repulsion.
\end{abstract}
\begin{document}

\flushbottom
\maketitle
%
%
\thispagestyle{empty}


\section*{Introduction}

The history of spin-triplet superconductivity dates back to the discovery of superfluidity for liquid $^3$He in 1972~\cite{Osheroff72}. After the discovery, its origin had been quickly resolved as a consequence of close collaboration between theory and experiment~\cite{Leggett75}. Since $^3$He is a fermion, the mechanism gives rise to the pair condensation into a macroscopic quantum state as required by superconductivity. An unusual feature of superfluid $^3$He is the role of internal degrees of freedom associated with $p$-wave spin-triplet ($S=1$) Cooper pair, i.e., combinations of spin states $|\uparrow\uparrow\rangle$, $(|\uparrow\downarrow\rangle+|\downarrow\uparrow\rangle)/\sqrt{2}$, $|\downarrow\downarrow\rangle$. This fact stimulated a long-standing challenge to discovery and understanding of spin-triplet superconductivity. Spin-triplet superconductors are being classified as {\it unconventional} in the sense that their physics are not described by the conventional $s$-wave spin-singlet ($S=0$) Bardeen-Cooper-Schrieffer (BCS) theory. In particular, the formation of Cooper pairs via Coulomb repulsion is still an important open issue in the field of unconventional superconductivity~\cite{Nandkishore12}.

In recent years, spin-triplet superconductivity has attracted much attention because of its intrinsic connection to spintronics and quantum computing applications~\cite{Eschrig11,Linder15}. The controlled production of spin-triplet supercurrents will open up perspectives for future novel superconducting devices. Pioneering theories predicted a generalization of spin-triplet components within ferromagnetic material proximity coupled to a spin-singlet superconductor~\cite{Bergeret01,Eschrig08}. This has been experimentally verified by observing, e.g., the emergence of odd-frequency spin-triplet states (pairs) at $s$-wave~\cite{Anwar10,Usman11,Srivastava17} and $d$-wave~\cite{Visani12,Kalcheim14,Kalcheim15,Komori18} superconductor/ferromagnetic interfaces. In conjunction with those observations, a direct penetration of {\it p}-wave spin-triplet pairs in a spin-triplet superconductor/ferromagnetic junctions~\cite{Anwar13,Anwar16} as well as an induction of {\it p}-wave superconductivity at {\it d}-wave superconductor/graphene interface~\cite{Bernardo17} are also very hot topics.

Practical studies regarding spin-triplet superconductivity was opened up by a synthesis of quasi-one-dimensional (Q1D) organic materials (TMTSF)$_2$X [X=PF$_6$, ClO$_4$], the so-called Bechgaard salts. In 1979, J\'erome {\it et al}. found a superconducting (SC) behavior of (TMTSF)$_2$PF$_6$ below the SC transition at $T_{\rm c}\sim20$ K~\cite{Jerome80}. A spin-density-wave insulating phase neighboring the SC phase suggests a significant role of electron correlation for the low-temperature physics~\cite{Jerome91}. In the early stage of nuclear magnetic resonance (NMR) measurements, spin-triplet superconductivity with a line-node gap was suggested for (TMTSF)$_2$PF$_6$~\cite{Takigawa87,Hasegawa87,Lee00,Lee01}. Later, however, a $^{77}$Se NMR Knight shift for (TMTSF)$_2$ClO$_4$ at low fields reveals a decrease in spin susceptibility $\chi_{\rm s}$ below $T_{\rm c}$, which indicates spin-singlet pairing~\cite{Shinagawa07}. Today, a $d$-wave spin-singlet superconductivity is most likely at low fields for (TMTSF)$_2$X; nevertheless, it has been suggested that there exists a phase transition or crossover to either a spin-triplet SC state~\cite{Shimahara00,Belmechri07} or an inhomogeneous Fulde-Ferrell-Larkin-Ovchinnikov state at high fields~\cite{Fulde64,Larkin65}. This scenario raises even more fascinating challenges in this field of research.

At present, the most promising material for spin-triplet superconductivity is the ruthenate Sr$_2$RuO$_4$. No sooner was discovered the SC transition at $T_{\rm c} \sim 1.5$ K~\cite{Maeno94}, the similarity between the superconductivity of Sr$_2$RuO$_4$ and the spin-triplet superfluidity of $^3$He was theoretically pointed out~\cite{Rice95}. After that, the spin-triplet pairing state having spontaneous time-reversal symmetry breaking with {\bf d}-vector perpendicular to the conducting plane was experimentally confirmed~\cite{Ishida98,Luke98}. Thus, analogous to the $^3$He-A phase, superconductivity in Sr$_2$RuO$_4$
has been generally concluded to be of the spin-triplet {\it p}-wave type~\cite{Rice98}.

Another candidate is water-intercalated sodium cobalt dioxides Na$_x$CoO$_2\cdot 1.3$H$_2$O ($x\sim0.35$), 
which exhibits a SC transition at $T_{\rm c} \approx 5$ K~\cite{Takada03}. By controlling $x$ using the trigonal CoO$_2$ distortion, two SC phases appear and they are separated by a narrow magnetic phase~\cite{Ohta10}. A theoretical multi-band tight-binding calculation speculated that the SC state for larger $x$ is of {\it s}-wave spin-singlet and that for smaller $x$ is of {\it p}- or {\it f}-wave spin-triplet. Experimentally, the relation between the two SC states as well as the origins of them are still highly controversial~\cite{Zheng06,Kobayashi08}

Very recently, a new family of Q1D superconductors A$_2$Cr$_3$As$_3$ (A=K, Rb, and Cs) has been recognized. At ambient pressure, a SC transition was observed at $T_{\rm c}=6.1$K for A=K~\cite{BaO15}, $4.8$K for A=Rb~\cite{Tang15a}, and $2.2$K for A=Cs~\cite{Tang15b}. It has been confirmed that the extrapolated upper critical field $H_{\rm c2}(T=0)$ largely exceeds the Pauli limit~\cite{BaO15,Tang15a,Tang15b}, which strongly supports spin-triplet pairings in the family compounds. Furthermore, the existence of nodes in the SC gap has been indicated by several experiments; linear behavior of the London penetration depth $\lambda(T)$ ~\cite{Pang15}, absence of the Hebel-Slichter coherence peak below $T_{\rm c}$ in the nuclear spin-lattice relaxation rate $1/T_1$~\cite{Zhi15,Yang15}, and Volovik-like field dependence ($\propto \sqrt{H}$) of the zero-temperature Sommerfeld coefficients $\gamma(H)$ in the SC mixed state~\cite{Tang15a}. From the theoretical aspect, spin-triplet Cooper pairs are considered to be formed simply by a ferromagnetic (FM) interaction~\cite{Fay80}. In fact, both the spin-lattice relaxation rate divided by temperature, $1/T_1T$, and the Knight shift $K$ increase when the temperature is decreased from $\sim100$K down to $T_{\rm c}$, suggesting significant FM spin fluctuations~\cite{Jiang15}.

Remarkably, the conduction pathway consists of coupled triangles in all the above four materials (see also Supplementary Information). It is known that a three-site Hubbard ring with two electrons generates a local FM interaction forming a triplet ground state rather than a singlet pairing~\cite{Tasaki98}. When these rings with pre-formed triplet pairs are coupled, a spin-triplet superconductivity might be expected just as the pre-formed singlets on rungs give rise to singlet-SC in ladder systems~\cite{Dagotto92,Balents96}. Previously, one of the present authors proposed a similar SC mechanism for a limited case of the TMTSF salts with anisotropic triangular structure including long-range Coulomb repulsions~\cite{Ohta05}. The aim of this paper is to construct a more general theory of spin-triplet SC mechanism for coupled odd-gons systems: a Hubbard ring with odd number of sites can provide a local FM interaction, and then spin-triplet superconductivity emerges through a melting of macroscopic ferromagnetism stabilized in the weakly coupled odd-gons systems. In this paper, we demonstrate the applicability of our SC mechanism by studying a twisted triangular Hubbard system as a model for Q1D superconductors A$_2$Cr$_3$As$_3$. The strong Q1D nature of this system enables us to perform very precise and unbiased numerical analyses using the density-matrix renormalization group (DMRG) method~\cite{White92}. Nevertheless, since the macroscopic ferromagnetism is expected to occur for any geometry of connections between the odd-gons (see below), this SC mechanism would be widely applicable to other real materials having crystal structure like triangular, kagome, pyrochlore lattices, and fullerenes, etc (see also Supplementary Information). It is also worth noting that our spin-triplet SC is mediated only by on-site Coulomb repulsion.

\begin{figure}
\centering
\includegraphics[width=1.0\linewidth]{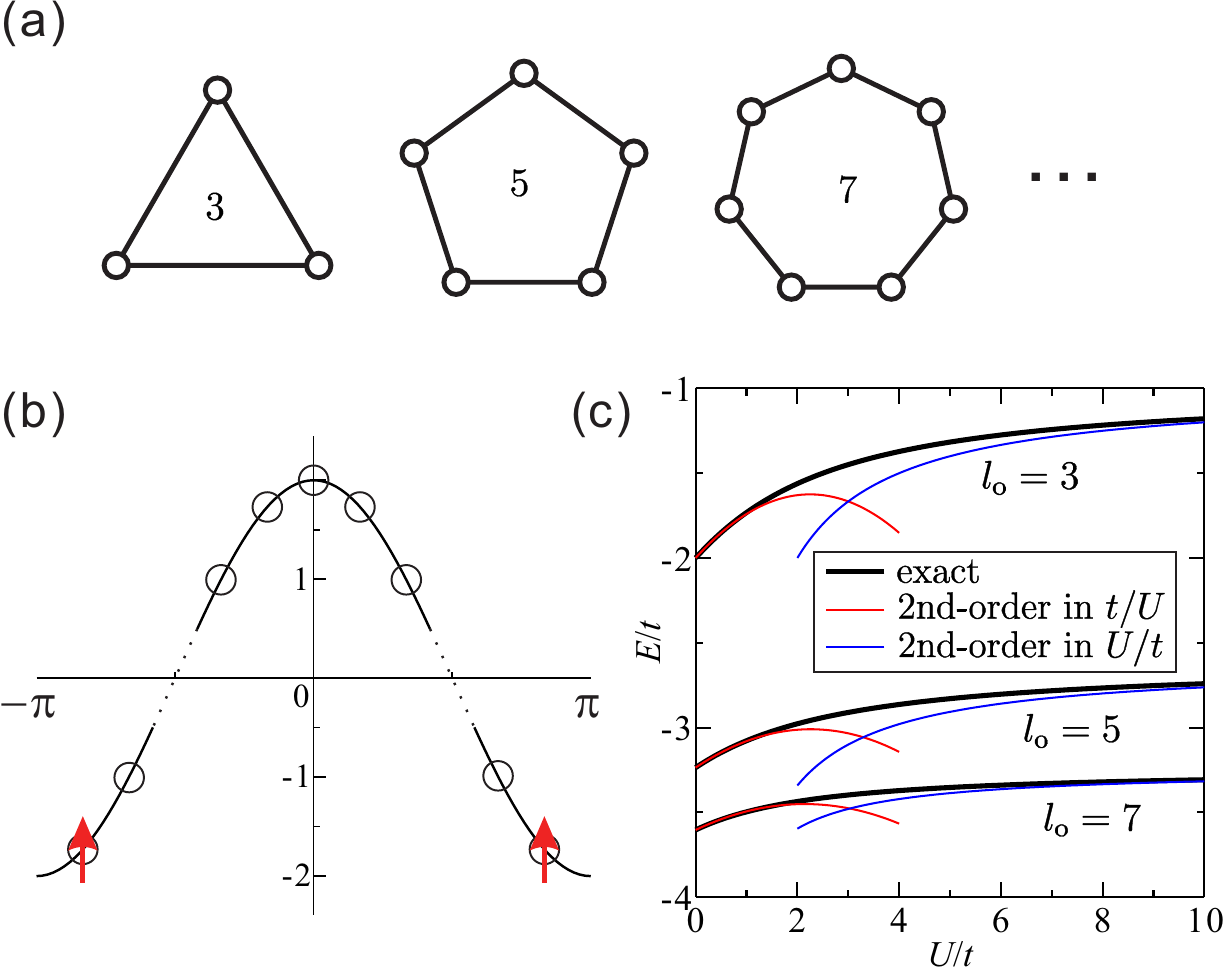}
\caption{
(a) Odd-gon clusters. (b) Dispersion of a single odd-gon, where the circles denote possible fermion shells. If the odd-gon contains two fermions, the lowest-lying shells are occupied by fermions with parallel spins in the ground state at $U>0$. (c) Energies of the spin-singlet state for $l_0=3$, $5$, and $7$. The weak- and strong-coupling second-order perturbation results are also shown (see Supplementary Information). Energies of the spin-triplet state are constant; -2, -3.2361, and -3.6039 for $l_0=3$, $5$, and $7$, which are equivalent to those of the spin-singlet state at $U=0$.
}
\label{fig_fm}
\end{figure}

\section*{Ferromagnetic mechanism}

Before performing the numerical analysis, we explain the FM mechanism which is generalized via a spin-triplet formation of two fermions in an isolated odd-gon Hubbard ring. We consider a Hubbard ring with odd number of sites $l_o$ containing two fermions [see Fig.~1(a)]. In momentum space the Hamiltonian is written as
\begin{eqnarray}
{\cal H}_{\rm ring}  &=&\sum_{k\sigma}\varepsilon (k) c^{\dagger}_{k\sigma}c_{k\sigma} + \frac{U}{l_o}\sum_{pkq}c^{\dagger}_{p-q\uparrow}c^{\dagger}_{k+q\downarrow}c_{k\downarrow}c_{p\uparrow}\\
\label{ham2}
\nonumber
&\equiv& {\cal H}_0+{\cal H}^\prime,
\end{eqnarray}
where $c_{k,\sigma}$ is an annihilation operator of fermion with spin $\sigma$ at momentum $k$ and $\varepsilon({k})=2t\cos k$ for uniform hopping integrals $t$. We assume $t>0$. The allowed momenta are $k=0, \pm2\pi/l_o, \pm4\pi/l_o, \cdots, \pm(l_o-1)\pi/l_o$. The energy dispersion is shown Fig.~\ref{fig_fm}(b).

When the spins of two fermions are parallel, namely in a spin-triplet state, the correlation term ${\cal H}^\prime$ vanishes. Thus, this spin-triplet state is exactly described as a two-fermion occupancy at $k=\pm(l_o-1)\pi/l_o$ levels [see Fig.~\ref{fig_fm}(b)] so that the total energy is $E_t=4t\cos\left(\frac{l_o-1}{l_o}\pi\right)$ independently of $U$. On the other hand, the correlation is involved in a spin-singlet state with anti-parallel spins. The total energy of the spin-singlet $E_s$ is plotted as a function of $U$ for $l_o=3$, $5$, and $7$ in Fig.~\ref{fig_fm}(c). We find that $E_t$ is always lower than $E_s$ for $U>0$. This immediately confirms the emergence of attractive interaction between two fermions in a spin-triplet state. Note that the total energy of a spin-triplet state with anti-parallel spins is also $E_t$ since the Hamiltonian (1) has the SU(2) symmetry.

We further consider the case of general $l_o$ by perturbation theory. In the weak-coupling limit ($U\ll1$) we take ${\cal H}^\prime$ as a perturbation to the unperturbed Hamiltonian ${\cal H}_0$. The unperturbed ground state ($U=0$) is exactly the same as the spin-triplet state; namely, the total energy is $E_t$. When the correlation is involved in a spin-singlet state the total energy is modified as
\begin{eqnarray}
E_s=E_t+\frac{U}{l_o}-\left(\frac{U}{l_o}\right)^2\left[\sin\left(\frac{l_o-1}{l_o}\pi\right)\sin\left(\frac{2}{l_o}\pi\right)\right]^{-1}
\label{2ndw}
\end{eqnarray}
up to the second order of $U$. Similarly, in the strong-coupling limit ($U\gg1$) the total energy is expressed as
\begin{eqnarray}
E=E_\infty-\frac{8}{Ul_o}\sin^2\left(\frac{2}{l_o}\pi\right)
\label{2nds}
\end{eqnarray}
up to the first order of $1/U$, where $E_\infty=4\sum_{n=3}^{l_o}\cos\left(\frac{n-1}{l_o}\pi\right)-2\cos\left(\frac{2}{l_o}\pi\right)$ is the total energy at $U=\infty$. Since the weak- and strong-coupling regimes are expected to be smoothly connected, a spin-triplet state would be always the ground state for any odd $l_o$ at $U>0$. The perturbation results (\ref{2ndw}) and (\ref{2nds}) are compared to the numerical results for $l_o=3$, $5$, and $7$ in Fig.~\ref{fig_fm}(c). This type of a spin-triplet formation has been frequently
discussed in the open-shell problem of a finite-size cluster. Here, the existence of (nearly-)degeneracy in the lowest-lying shells is essential to obtain a spin-triplet ground state; and, such the situation is easily realized for odd $l_o$ and $U, t>0$. So, the question we address here is whether a triplet SC is stabilized when these spin-triplet odd-gon units are hybridized.

\begin{figure}
\centering
\includegraphics[width=1.0\linewidth]{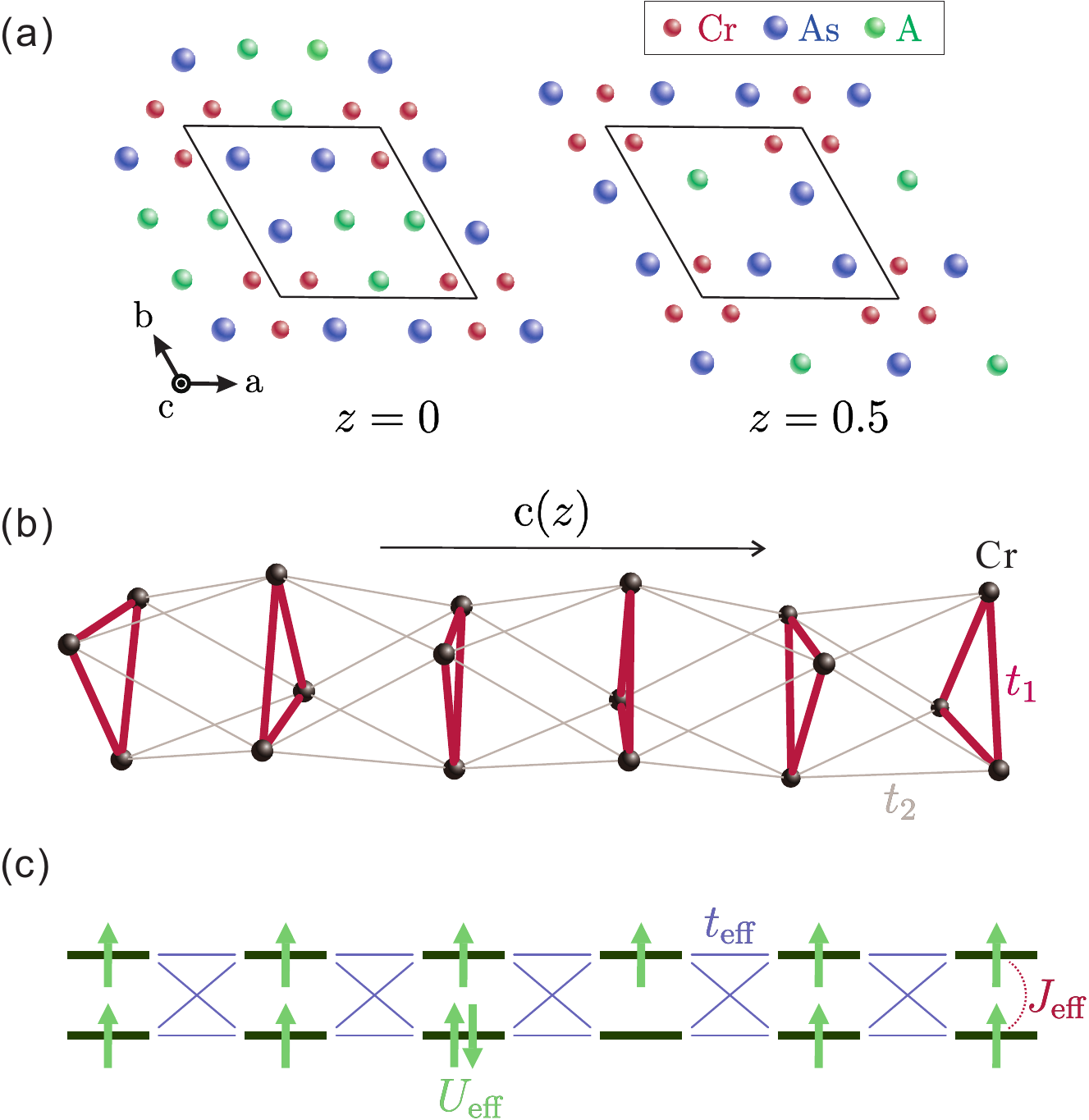}
\caption{
(a) Crystal structure of A$_2$Cr$_3$As$_3$ for two alternating $ab$-planes. (b) Lattice structure of the twisted triangular tube, as a model of A$_2$Cr$_3$As$_3$. (c) A simplified effective model for K$_2$Cr$_3$As$_3$.
}
\label{fig_lattice}
\end{figure}

\section*{Model}

Let us then couple the odd-gon units. A good playground with coupled odd-gon units is provided by the chromium arsenide A$_2$Cr$_3$As$_3$. The crystal structure consists of double-walled subnanotubes [Cr$_3$As$_3$]$_2$ [see Fig.~\ref{fig_lattice}(a)]. They form Q1D chains by stacking of triangular units along the $c$-axis and these Q1D chains are separated by columns of A$^+$ ions. First-principles calculations~\cite{Jiang15,Alemany15} found that the conductivity is dominated by the $d_{z^2}$, $d_{x^2-y^2}$, and $d_{xy}$
orbitals of Cr atom, and three energy bands are involved at the Fermi level; two Q1D and a 3D. Besides, experimentally, the strong 1D nature of conduction along the $c$-axis has been confirmed by detecting the Tomonaga-Luttinger liquid behaviors~\cite{Zhi15,Watson17}. Therefore, it would be reasonable to consider the so-called ``twisted triangular Hubbard tube'' as a model to study the conduction properties of A$_2$Cr$_3$As$_3$ by taking the highest occupied molecular orbital (HOMO) in Cr to be a Hubbard site~\cite{Zhong15} [see Fig.~\ref{fig_lattice}(b)]. The Hamiltonian in real space is
\begin{equation}
{\cal H}=\sum_{i,j,\sigma}t_{ij}(c_{i,\sigma}^\dag c_{j,\sigma} + h.c.)+U\sum_i n_{i\uparrow} n_{i\downarrow},
\label{ham1}
\end{equation}
where $c_{i,\sigma}$ is an annihilation operator of electron with spin $\sigma$ at site $i$, $t_{ij}$ is hopping integral between sites $i$ and $j$, $U(>0)$ is on-site Coulomb interaction, and $n_{k\sigma}=c^\dagger_{i\uparrow}c_{i\downarrow}$ . Using the Slater-Koster tight-binding method, the sign of intra-triangle hopping integral $t_1$ is estimated to be positive in hole notation (see Supplementary Information). This meets the FM condition introduced in the previous section. We take $t_1=1$ as the unit of energy hereafter. Note that the system is gauge-invariant under a transformation of inter-triangle hopping integral $t_2 \to -t_2$. We use $U=10$ as a typical value for strongly correlated electron systems.

\section*{Numerical calculations}

\begin{figure}
\centering
\includegraphics[width=1.0\linewidth]{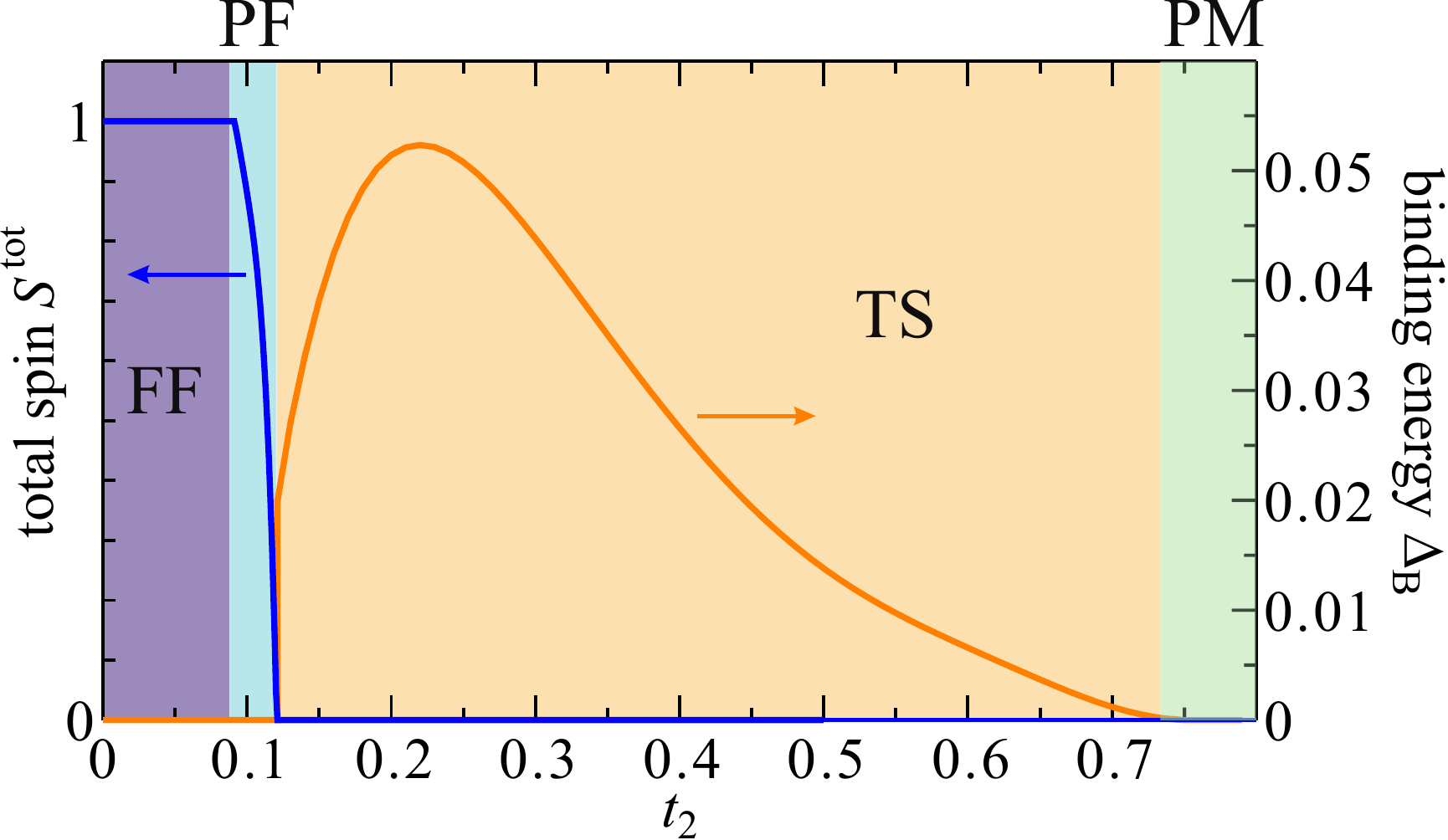}
\caption{
Normalized total spin $S^{\rm tot}$ and binding energy $\Delta_{\rm B}$ at $n=\frac{1}{2}$. The results are plotted as a function of $t_2$.
}
\label{fig_transition}
\end{figure}

We now present the DMRG results for the twisted triangular Hubbard tube. We studied clusters up to $L\times3=60\times3$ sites under the open boundary conditions with keeping $m=1000$ to $12000$ density-matrix eigenstates. All the physical quantities have been extrapolated to the $m \to \infty$ and $L \to \infty$ limits. First, we consider the system at $n=2/3$ filling where the spin-triplet ($S=1$) triangle units are coupled by $t_2$ along the tube direction. It is found the the system remains insulating with large charge gap and can be treated as $S=1$ spin chain system~\cite{Janani14}. The situation drastically changes away from $n=2/3$ filling. Let us consider the case of a smaller filling $n=1/2$. First, we calculate the total spin $S^{\rm tot}$ defined as $\langle \vec{S}^2 \rangle = \sum_{i,j} \langle \vec{S}_i \cdot \vec{S}_j \rangle = S^{\rm tot}(S^{\rm tot}+1)$. In Fig.~\ref{fig_transition} the $L\to\infty$ extrapolated result of total spin per site is plotted as a function of $t_2$. Quite interestingly, an infinitesimal $t_2$ induces a fully-polarized FM (FF) order even though there is no explicit FM interactions between the triangles. {\it This implies that the geometry of inter-triangle network is not essential for the appearance of the FF state}; in other words, the spins are aligned only if hole(s) can hop from odd-gon to odd-gon. This may be similar to the Nagaoka mechanism~\cite{Nagaoka66} in the sense that hole propagation stabilizes a ferromagnetism. The dependence of the FM stabilization on the on-site Coulomb interaction $U$ is demonstrated in Supplementary Information. With increasing $t_2$, the FF order starts to melt at $t_2=0.092$ and the system goes into a narrow partially-polarized FM (PF) phase at $0.092<t_2<0.121$. The FM polarization completely vanishes at $t_2=0.121$ and a global singlet state characterized by $S^{\rm tot}=0$ follows. Nonetheless, the short-range (at least within the odd-gon units) spin-spin correlation is naturally expected to be FM even after the melting of FM order.

\begin{figure}
\centering
\includegraphics[width=1.0\linewidth]{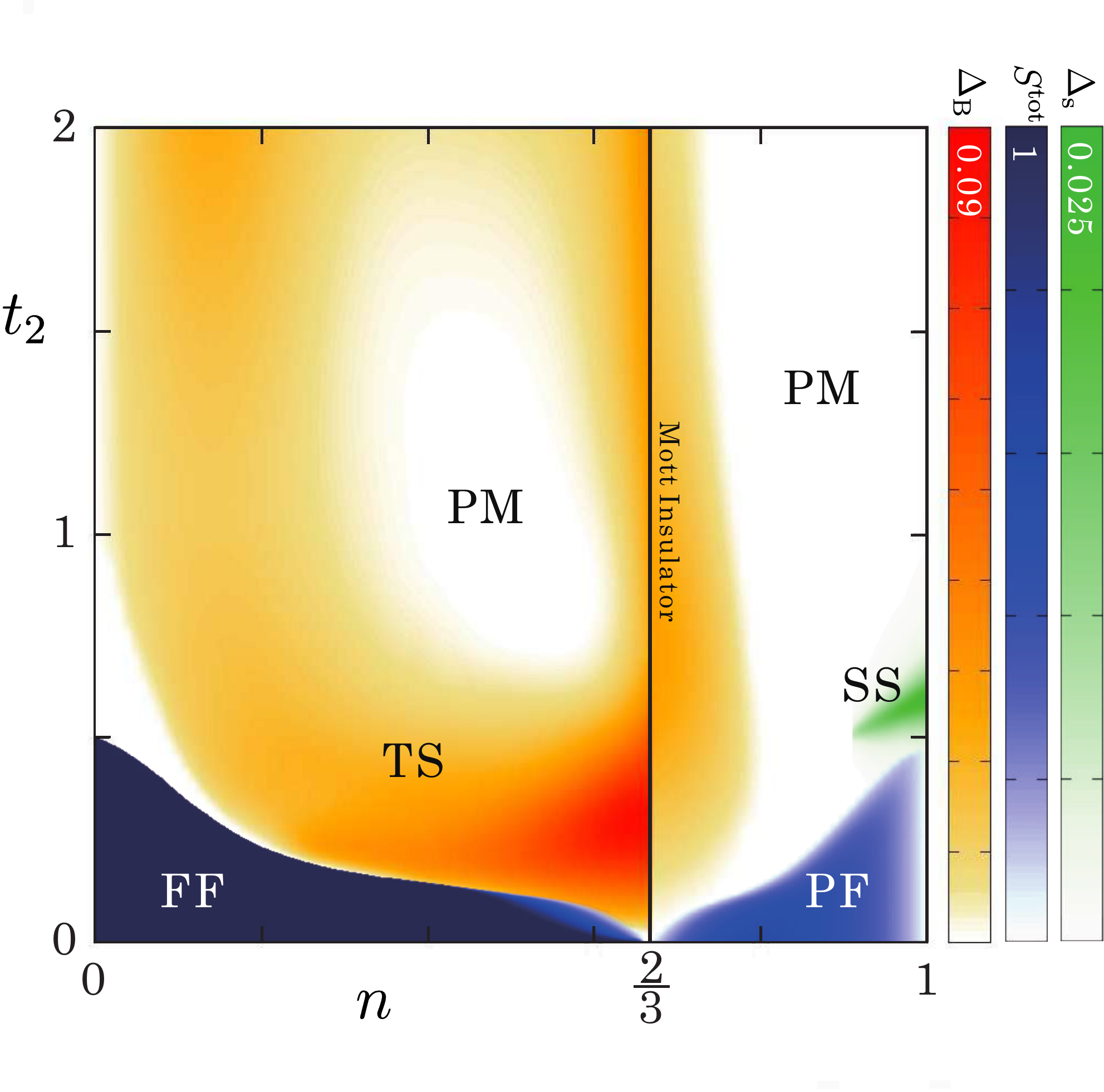}
\caption{
Ground-state phase diagram of the twisted triangular Hubbard tube at $U=10$. The values of binding energy $\Delta_{\rm B}$, normalized total spin $S^{\rm tot}$, and spin gap $\Delta_{\rm s}$ are indicated by the shading.
Different six phases are included (FF: fully-polarized ferromagnetic phase, PF: partially-polarized ferromagnetic phase, PM: paramagnetic metallic phase, TS: spin-triplet superconducting phase, SS: spin-singlet superconducting phase, and the Mott insulating phase at $n=\frac{2}{3}$).
}
\label{fig_pd}
\end{figure}

Thus, the next question is whether the short-range FM correlation stabilizes spin-triplet bound pairs of fermions. We determine this by examining binding energy of fermions, defined as $\Delta_{\rm B}=-\lim_{L\to\infty}[E_0(N_\uparrow\pm1,N_\downarrow\pm1;L)+E_0(N_\uparrow,N_\downarrow;L)-2E_0(N_\uparrow\pm1,N_\downarrow;L)]$, where $E_0(N_\uparrow,N_\downarrow;L)$ is the ground-state energy of the system with length $L$ containing $N_\uparrow$ spin-up and $N_\downarrow$ spin-down fermions. The $L\to\infty$ extrapolated result of $\Delta_{\rm B}$
is plotted in Fig.~\ref{fig_transition}. The finite-size scaling analysis is illustrated in Supplementary Information. A finite binding energy in the thermodynamic limit indicates an effective attractive interaction of the sort necessary to mediate superconducting pairing. We see that $\Delta_{\rm B}$ which jumps from $0$ to $0.0199$ at $t_2=0.121$, indicating a first-order transition, goes through a maximum at $t_2\sim0.22$, and decreases due to an increase of geometrical frustration (see below). The system goes into a paramagnetic (PM) metallic state ($\Delta_{\rm B}=0$) at $t_2\sim0.732$. We have also confirmed that the charge gap, defined as
$\Delta_{\rm c}=\lim_{L\to\infty}[E_0(N_\uparrow+1,N_\downarrow+1;L)+E_0(N_\uparrow-1;N_\downarrow-1;L)-2E_L(N_\uparrow,N_\downarrow)]/2$,
and the spin gap, defined as $\Delta_{\rm s}=\lim_{L\to\infty}[E_0(N_\uparrow+1,N_\downarrow-1;L)-E_L(N_\uparrow,N_\downarrow)]$, are always zero for all $t_2$ at $n=1/2$. Therefore, the region with $\Delta_{\rm B}>0$ is characterized as a spin-triplet SC phase. The same quantities are also calculated for $0<n<1$ and the results are summarized as a phase diagram in Fig.~\ref{fig_pd}. We found a wide region of the spin-triplet SC phase and the binding energy is most enhanced around $n=\frac{2}{3}$. In fact, the Q1D HOMO band of A$_2$Cr$_3$As$_3$
is nearly $\frac{2}{3}$-filled with holes (or $\frac{4}{3}$-filled with electrons)~\cite{Jiang15}. For further confirmation of the opening of SC gap, it would be interesting to calculate the anomalous Green’s functions for Bogoliubov quasiparticle excitations~\cite{Ohta94} in future.

\section*{Effective model}

In this section, an effective model is constructed for the twisted triangular Hubbard tube to explain the phase diagram [Fig.~\ref{fig_pd}] more intuitively. Here we illustrate its construction for a coupled triangular unit system but a similar consideration is possible for any coupled odd-gon units system. Each of the triangles can be mapped onto a site with two orbitals at $0<n<1$ which correspond to the two lowest-lying shells in the momentum space [see Fig.~\ref{fig_fm}(b)]. The intra-triangle FM interaction is captured as Hund's coupling $J_{\rm eff} (<0)$ between the intra-site orbitals. The amount is given by the energy difference between the spin-triplet ground state and first spin-singlet excited state of the isolated triangle including two fermions, i.e., $J_{\rm eff}=E_s-E_t$ [see Fig.~\ref{fig_fm}(c)]. Further, the on-site Coulomb repulsion $U_{\rm eff}$, which is the origin of Mott state at $n=\frac{2}{3}$, is approximately estimated by the single-particle gap of the isolated triangle. The estimation at $n=\frac{2}{3}$
is given in Supplementary Information. The inter-site hopping parameter $t_{\rm eff}$ is simply proportional to $t_2$. A schematic picture of our effective model is sketched in Fig.~\ref{fig_lattice}(c).

We now give an interpretation of the phase diagram using the effective model. Let us start in the small $t_{\rm eff}$ limit. At $n=\frac{2}{3}$ the system is insulating because two `parallel' spins are confined in each of the sites. The degrees of freedom of each the site is spin-1. An AFM interaction between the inter-site orbitals is induced by nonzero $t_{\rm eff}$ and the magnetism is described by the Affleck-Kennedy-Lieb-Tasaki state in the spin-1 Heisenberg chain~\cite{Affleck87}. Surprisingly, the ground state is drastically changed when the system is doped by fermions or holes. In either case the spins polarize to avoid feeling $U_{\rm eff}$ and also to gain $J_{\rm eff}$ in the inter-triangle hopping processes. It is similar to the Nagaoka ferromagnetism~\cite{Nagaoka66} in the sense that the spins are ferromagnetically aligned due to the movement of particles in the system. Connecting with infinitesimally small $t_{\rm eff}$, Therefore, the total spin is $\frac{N}{2}\left(\frac{N}{2}+1\right)$ at $n<\frac{2}{3}$
and $\frac{4L-N}{2}\left(\frac{4L-N}{2}+1\right)$ at $n>\frac{2}{3}$.
In the latter case the spins are partially screened because more than two electrons on average are contained in a site. Now consider what happens when we increase $t_{\rm eff} (\propto t_2)$. With increasing $t_{\rm eff}$, a collapse of the FM polarization is expected since the AFM interaction between the inter-site orbitals increases $(\propto \frac{t_{\rm eff}^2}{U_{\rm eff}})$. In fact, the both FF and PF phases disappear at larger $t_{\rm eff}$ for all $n$. Especially, the FM polarization is immediately destroyed by small $t_{\rm eff}$ near $n=\frac{2}{3}$ where the AFM interaction is maximized like in the half-filled Hubbard model.

As $t_{\rm eff}$ increases, the FF and PF phases depending on electronic fillings are followed by a global singlet ($S^{\rm tot}=0$) phases; spin-triplet SC (TS), PM, and spin-singlet SC (SS) phases. The appearance of the TS phase may be rather naturally expected because the increase of $t_{\rm eff}$ enhances only the inter-site AFM interaction and does not affect the on-site triplet pairings. However, the SS phase can not be explained within our effective model because it does not include an AFM interaction with the triangular unit to generate singlet pairs.
But the coupled triangular units system at $n=1$ is magnetically frustrated and in some cases the ground state is spontaneously dimerized, namely, the spin-singlet pairs are locally formed to relax the magnetic frustration~\cite{Kawano97,Fouet06}. Thus, our SS state may be interpreted as a doped valence-bond-solid state. So, the PM would be just a crossover between TS and SS phases. They are mostly distributed near $n=1$. The minimized binding energy and the appearance of PM near $t_1=t_2$, where the frustration is maximal, at $n<\frac{2}{3}$ can be also explained in the same sense. The binding energy goes up again with increasing $t_2$ for $t_2 \ge t_1$.

\begin{figure}
\centering
\includegraphics[width=1.0\linewidth]{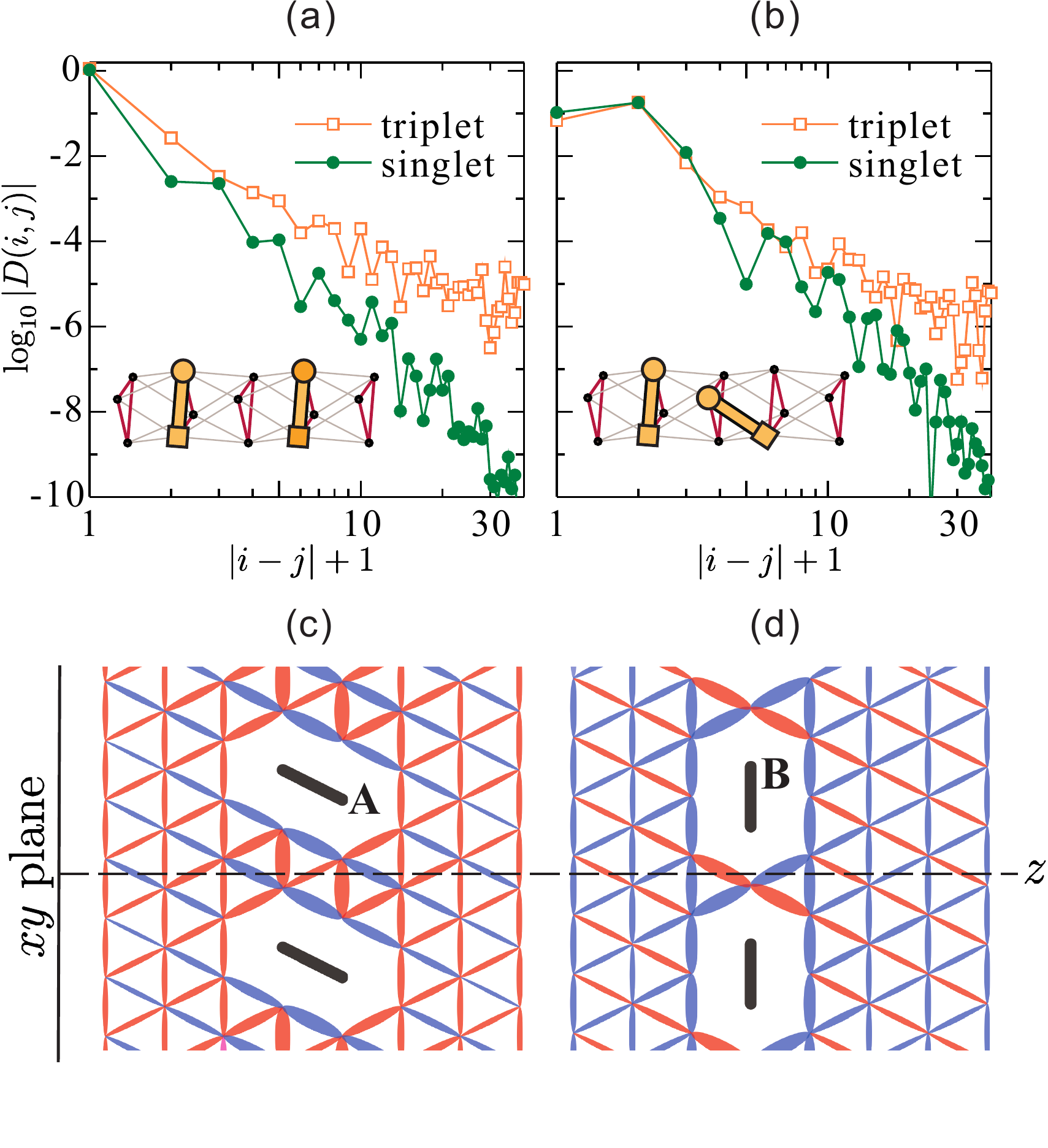}
\caption{
Pair-field correlation functions $D(i,j)$ between (a) two intra-triangle bonds and (b) intra-triangle and inter-triangle bonds, calculated with $L=60\times3$ cluster for $t_2=0.25$ at $n=1/2$. In the inset, the circle and square denote the sites $\lambda=1$ and $\lambda=2$ within a pair bond, respectively. Horizontal axis $|i-j|+1$ counts the distance between the circles along the conduction ($z$) direction. (c)(d) Spin-triplet pair-field correlation function of the twisted triangular Hubbard tube, where the twisted triangular tube is expanded along the side of $t_1$ triangles parallel to the $xy$-plane and the conducting $c$ direction is taken to be the $z$-axis. The reference bonds {\bf A} and {\bf B} are indicated by thick-black lines. The width of each ellipse for a bond at position $i$ is proportional to $1/\log|D(i,{\bf A})|$ and  $1/\log|D(i,{\bf B})|$ in (c) and (d), respectively. Blue (red) color represents positive (negative) value of the correlation function.
}
\label{fig_SCcorr}
\end{figure}

\section*{Pairing symmetry}

Finally, we discuss the pairing symmetry of our twisted triangular Hubbard model. So far, the possibilities of {\it f}-wave and {\it p}$_z$-wave symmetries have been suggested both theoretically and experimentally for the superconductivity in K$_2$Cr$_3$As$_3$. To determine it numerically, we calculated the pair-field correlation function $D(i,j)=\langle \Delta_i^\dagger \Delta_j \rangle$ with $\Delta_i=c_{i1\uparrow} c_{i2\downarrow}-c_{i1\downarrow} c_{i2\uparrow}$ for singlet pairs and
$\Delta_i=c_{i1\uparrow} c_{i2\downarrow}+c_{i1\downarrow} c_{i2\uparrow}$ for triplet pairs. Here, $c_{i\lambda\sigma}$ annihilates an electron of spin $\sigma$ on a site $\lambda$ ($=1$ or $2$) of the bond indexed by $i$. For reference, the correlation function $|D(i,j)|$ is plotted on a log-log scale as a function of distance $|i-j|-1$ in Fig.~\ref{fig_SCcorr}(a)(b). Although the system length is too short to estimate the decay ratio accurately, the spin-triplet pair-field correlation is obviously much more dominant than the spin-singlet one. Let us then see the sign distribution of the spin-triplet pair-field correlation function. The results are shown in Fig.~\ref{fig_SCcorr}(c)(d), where the twisted triangular tube is expanded along the side of $t_1$ triangles parallel to the $xy$-plane (crystallographic {\it ab}-plane) and the conducting {\it c}-direction is taken to be the $z$-axis. Positive correlations are represented by blue color and negative correlations are represented by red color. We found that the correlation function keeps the sign unchanged for a rotation about the $z$-axis, namely, all $D(i,j)$'s parallel to the $xy$ plane have the same sign. This means that the pairing symmetry is isotropic with a rotation about the $z$-axis. While, $D(i,j)$ changes its sign alternately along the $z$ line, e.g., on a dotted line in Fig.~5(c)(d). This is consistent with the {\it p}$_z$-wave pairing symmetry. More details are given in Supplementary Information.

\section*{Conclusion and discussion}

We proposed a universal mechanism for spin-triplet superconductivity (SC) in a coupled odd-gons (e.g., triangular unit) Hubbard system. First, we show that two fermions on a Hubbard ring with odd number of sites form a spin-triplet pair in the ground state. When the odd-gons are weakly coupled by hopping integral, a global FM order is induced by particle moving between the odd-gons; and with increasing the hopping integral, a spin-triplet SC state appears through a melting of the FM order. We demonstrated the validity of this mechanism by considering the twisted triangular Hubbard tube as a model of Q1D superconductors A$_2$Cr$_3$As$_3$ using the DMRG technique. From the analogy of the high-temperature SC which is located next to the AFM insulating phase, it is interesting that the SC pairing energy is most enhanced in the vicinity of the Mott-Hubbard metal-to-insulator transition at filling $n=2/l_o$. We then derived a simple effective model, namely, two-orbital Hubbard model with inter-orbital FM interaction, to provide a general application to coupled odd-gons system. We also confirmed that the spin-triplet pairing of the twisted triangular Hubbard tube occurs predominantly in the {\it p}$_z$-wave channel.

The Nagaoka~\cite{Nagaoka66} and flat-band~\cite{Mielke91} mechanisms are well known as the origin of ferromagnetism. However, both of them are unsuitable for explaining the spin-triplet SC since they lead not to Cooper pairs but only to a saturated magnetization. On the other hand, in our model, a global ferromagnetism is generated which melts away to triplet SC when the local spin-triplet pairs in odd-gons are coupled. It is similar to the situation where preformed singlet pairs on rungs give rise to singlet SC when these are coupled to form a two-leg ladder system. 
Since the geometry of hopping network between the odd-gons is not essential for the appearance of global ferromagnetism, there could be many candidates for spin-triplet SC as coupled odd-gons system, for example, kagome systems (coupled triangles), pyrochlore systems (coupled triangles), fullerenes (coupled pentagons), and vanadium oxide Na$_2$V$_3$O$_7$ (coupled enneagons i.e., polygon with 9 sides), etc. We thus argue that the materials consisting of odd-numbered geometric units would be a treasure house of spin-triplet SC. We hope that our study could widely open up the opportunities to find spin-triplet superconductivity.


\section*{Acknowledgements}

This work is supported by SFB 1143 of the Deutsche Forschungsgemeinschaft and the NSF through Grants
No. DMR-1506263 and No. DMR-1506460. Computations were carried out on the ITF/IFW computer cluster. We thank U. Nitzsche for technical assistance.

\section*{Author contributions statement}

S.N. designed the study. All authors performed numerical calculations and analyzed the results. S.N. drafted the manuscript. All authors approved the final manuscript.

\section*{Additional information}


\textbf{Competing financial interests} The authors declare no competing interests.

\clearpage

\large
\begin{center}
{\bf Supplemental Information for\\
Triplet superconductivity in coupled odd-gon rings}
\end{center}

\normalsize

\renewcommand{\theequation}{S\arabic{equation}}
\renewcommand{\thefigure}{S\arabic{figure}}
\renewcommand{\thetable}{S\arabic{table}}
\setcounter{equation}{0}
\setcounter{figure}{0}
\setcounter{table}{0}

\section{Transfer integrals for A$_2$Cr$_3$As$_3$ (A=K, Rb, and Cs)}

\begin{figure}[ht]
\centering
\includegraphics[width=0.6\linewidth]{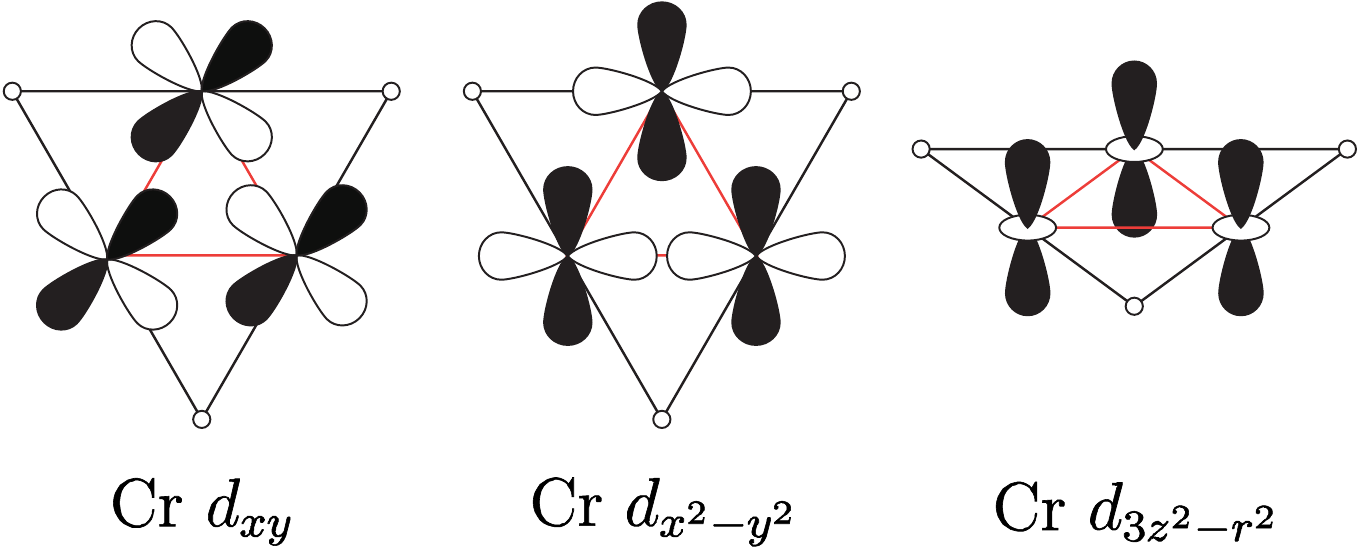}
\caption{
Schematic pictures of chromium 3$d$ orbitals dominating the conductivity in A$_2$Cr$_3$As$_3$.
}
\label{fig_orbitals}
\end{figure}

In the main text, a model for A$_2$Cr$_3$As$_3$ was constructed by taking a Cr site as a Hubbard site. Therefore, we here estimate the transfer integrals between the Cr sites within the Slater-Koster (SK) parameterization~\cite{Slater54}. As shown in Fig.~\ref{fig_orbitals}, the intra-triangle ($ab$ plane) conductivity is mostly generated from the direct hybridizations between Cu $d_{xy}$ orbitals:
\begin{equation}
E_{xy,xy}=3l^2m^2V_{dd\sigma}+(l^2+m^2-4l^2m^2)V_{dd\pi}+(n^2+l^2m^2)V_{dd\delta},
\end{equation}
and between Cu $d_{x^2-y^2}$ orbitals:
\begin{equation}
E_{x^2-y^2,x^2-y^2}=\frac{3}{4}(l^2-m^2)^2V_{dd\sigma}+[l^2+m^2-(l^2-m^2)^2]V_{dd\pi},
\end{equation}
where $V_{dd\sigma}$, $V_{dd\pi}$, and $V_{dd\delta}$ are the bond integrals for $\sigma$, $\pi$, and $\delta$ bonds, respectively. The interatomic vector is expressed as $\vec{r}_{i,j}=(r_x,r_y,r_z)=d(l,m,n)$, where $d$ is the distance between the atoms and $l$, $m$, and $n$ are the direction cosines to the neighboring atom. By the Muffin-Tin Orbital theory and pseudopotential theory, the bond integral is obtained as
\begin{equation}
V_{ddn}=\eta_{ddn}\frac{\hbar r_d^3}{m d^5},
\end{equation}
where $r_d$ is a characteristic length of transition metal; it is 0.90\AA for Cr, and $\eta_{dd\sigma}=-\frac{45}{\pi}$, $\eta_{dd\pi}=\frac{30}{\pi}$, and $\eta_{dd\delta}=-\frac{15}{2\pi}$~\cite{Harrison80}. Using the crystal structure determined by the X-ray diffraction~\cite{Bao15}, we obtain
\begin{equation}
E_{xy,xy}+E_{x^2-y^2,x^2-y^2}=-1.194\frac{\hbar r_d^3}{m d^5}
\end{equation}
in electron notation, namely, the intra-triangle transfer integral in hole notation is $1.194\frac{\hbar r_d^3}{m d^5}>0$. This corresponds to $t_1$ in the main text. Whereas, the transfer integrals along the $c$-axis, $t_2$, is estimated by the hybridization between Cr $d_{3z^2-r^2}$ orbitals:
\begin{equation}
E_{3z^2-r^2,3z^2-r^2}=[n^2-\frac{1}{2}(l^2+m^2)]^2V_{dd\sigma}+3n^2(l^2+m^2)V_{dd\pi}+\frac{3}{4}(l^2+m^2)^2V_{dd\delta}
\end{equation}
This value may be a few times larger than $t_1$. So we present the phase diagram for $t_2=0$ to $2t_1$ in the
main text.

\section{Stabilization of ferromagnetism on the on-site Coulomb repulsion}

\begin{figure}[htb]
\centering
\includegraphics[width=0.5\linewidth]{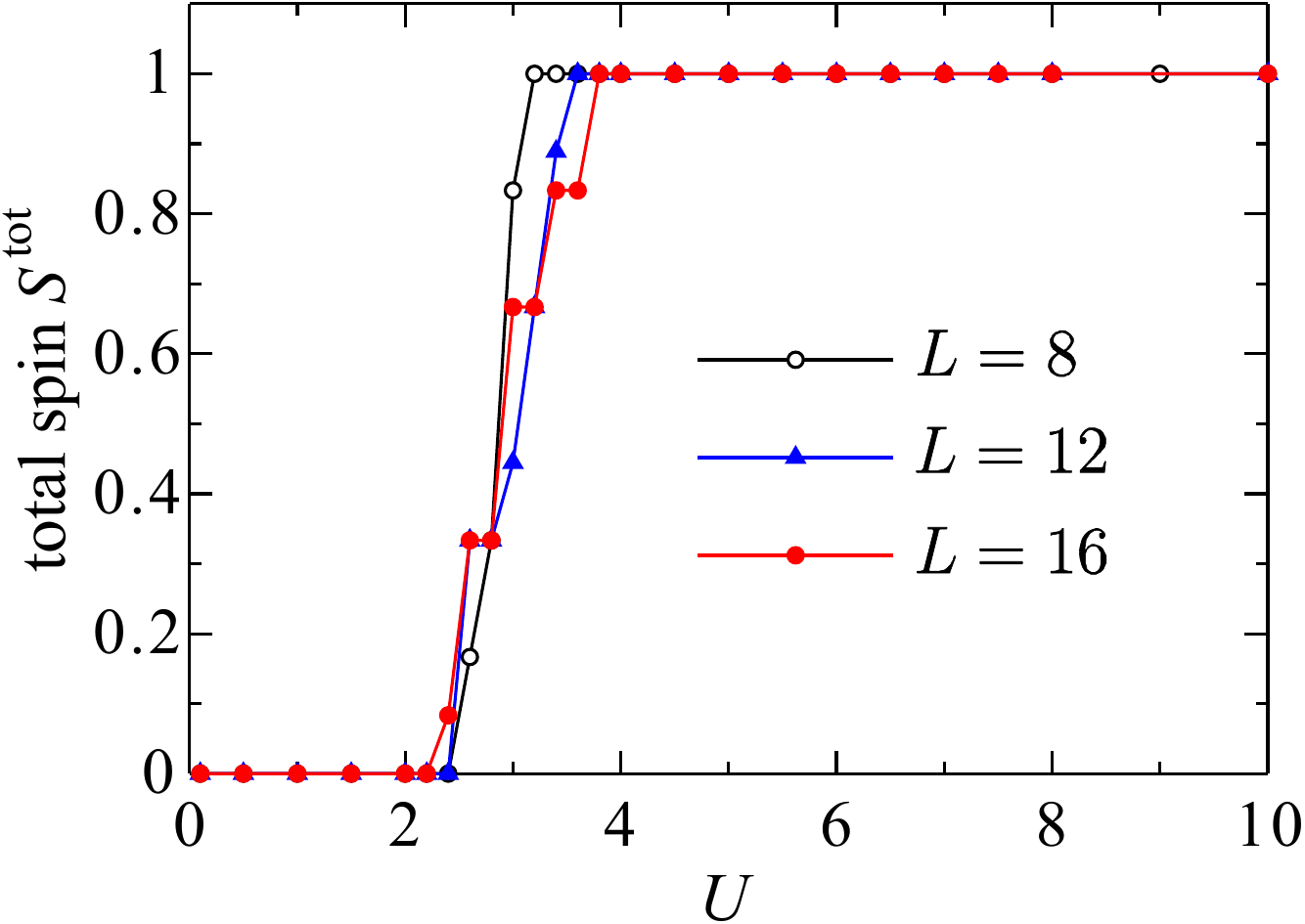}
\caption{
Renormalized total spin as a function of the on-site Coulomb interaction for the twisted Hubbard model at $t_2=0.05$ and $n=1/2$ for different clusters $L \times 3$.
}
\label{fig_Stot_U}
\end{figure}

To illustrate the effect of on-site Coulomb interaction $U$ on the global ferromagnetic ordering in the twisted Hubbard model, we show the total spin as a function of $U$ at $t_2=0.05$ and $n=1/2$ in Fig.~\ref{fig_Stot_U}. We can see that the fully polarized state is stabilized for larger $U$, where the ferromagnetic interaction on each odd-gon, namely, $J_{\rm eff}$ in our effective model, is stronger. The ferromagnetic interaction is estimated from the energy difference between the spin-triplet ground state and first spin-singlet excited state of the isolated triangle including two fermions. This interaction saturates quickly with increasing $U$, as seen in Fig.~1 (c) of the main text, and the global ferromagnetic ordering can be stabilized even by relatively small $U$.

\section{Coulomb repulsion in the effective model}

In the main text we introduced an effective model to describe the ferromagnetism and spin-triplet superconductivity of the twisted triangular Hubbard tube. The Coulomb repulsion works when the effective site including two orbitals is occupied by three fermions. The Coulomb repulsion is roughly estimated from the single-particle gap of the isolated odd-gon, namely, $U_{\rm eff}=(E_3-E_2)-(E_2-E_1)$ where $E_N$ is the ground-state energy of the odd-gon Hubbard ring with $N$ fermions. For example, in the case of $l_o=3$, $E_1=-t$, $E_2=-2t$, and $E_3$ can be obtained by solving an equation $E_3^3-2UE_3^2+(U^2-9t^2)E_3+6Ut^2=0$ where $U$ and $t$ are the on-site Coulomb interaction and hopping integral in the original triangle.

\section{Conductive networks of (TMTSF)$_2$X, Sr$_2$RuO$_4$ and Na$_{0.35}$CoO$_2$$\cdot$$1.3$H$_2$O}

\subsection{(TMTSF)$_2$X}

The crystal structure of the Bechgaard salts (TMTSF)$_2$X consists of well-separated sheets containing one-dimensional TMTSF stacks along the $a$-axis. The sheets are in the $ab$-plane and the transfer integrals along the $b$-axis are about 10 - 20$\%$ of those along the $a$-axis~\cite{Ducasse86}. The unique structure of the transfer integrals can be regarded as an anisotropic triangular lattice. There are three electrons in the two highest-occupied molecular orbitals of a dimerized molecules, e.g., (TMTSF)$_2$, and the system is at $\frac{3}{4}$-filling in terms of electrons, which corresponds to $\frac{1}{4}$-filling in terms of holes~\cite{Jerome94}.

\subsection{Sr$_2$RuO$_4$}

The ruthenate Sr$_2$RuO$_4$ is a tetragonal, layered perovskite system of stacking RuO$_2$-planes. Like Cu in the high-$T_{\rm c}$ superconductors, Ru atoms form a square lattice. If a single-band description of RuO$_2$-plane for so-called $\gamma$ band could be adequate, the system is described as a 2D Hubbard model with next-nearest-neighbor hopping. The next-nearest-neighbor transfer integral has been estimated to be $0.3 - 0.4$ in units of the nearest-neighbor transfer integral~\cite{Mazin97,Liebsch00}, so that Ru indeed forms a triangular network. Based on the quantum  oscillation  measurement, the $\gamma$ Fermi surface sheet is a large electron-like cylinder with the electron filling $n \sim 2/3$.

\subsection{Na$_{0.35}$CoO$_2$$\cdot$$1.3$H$_2$O}

In the cobalt oxide Na$_{0.35}$CoO$_2$$\cdot$$1.3$H$_2$O, the conductive $ab$ planes consist of edge-sharing CoO$_6$ octahedra and each plane is strongly separated by Na$^+$ ions and H$_2$O molecules along $c$ axis. The Co ions form a triangular lattice, and the system may be regarded as a two-dimensional triangular lattice doped with $35\%$ electrons.

\section{2D lattice as a coupled odd-gons}

\begin{figure}[t]
\centering
\includegraphics[width=0.6\linewidth]{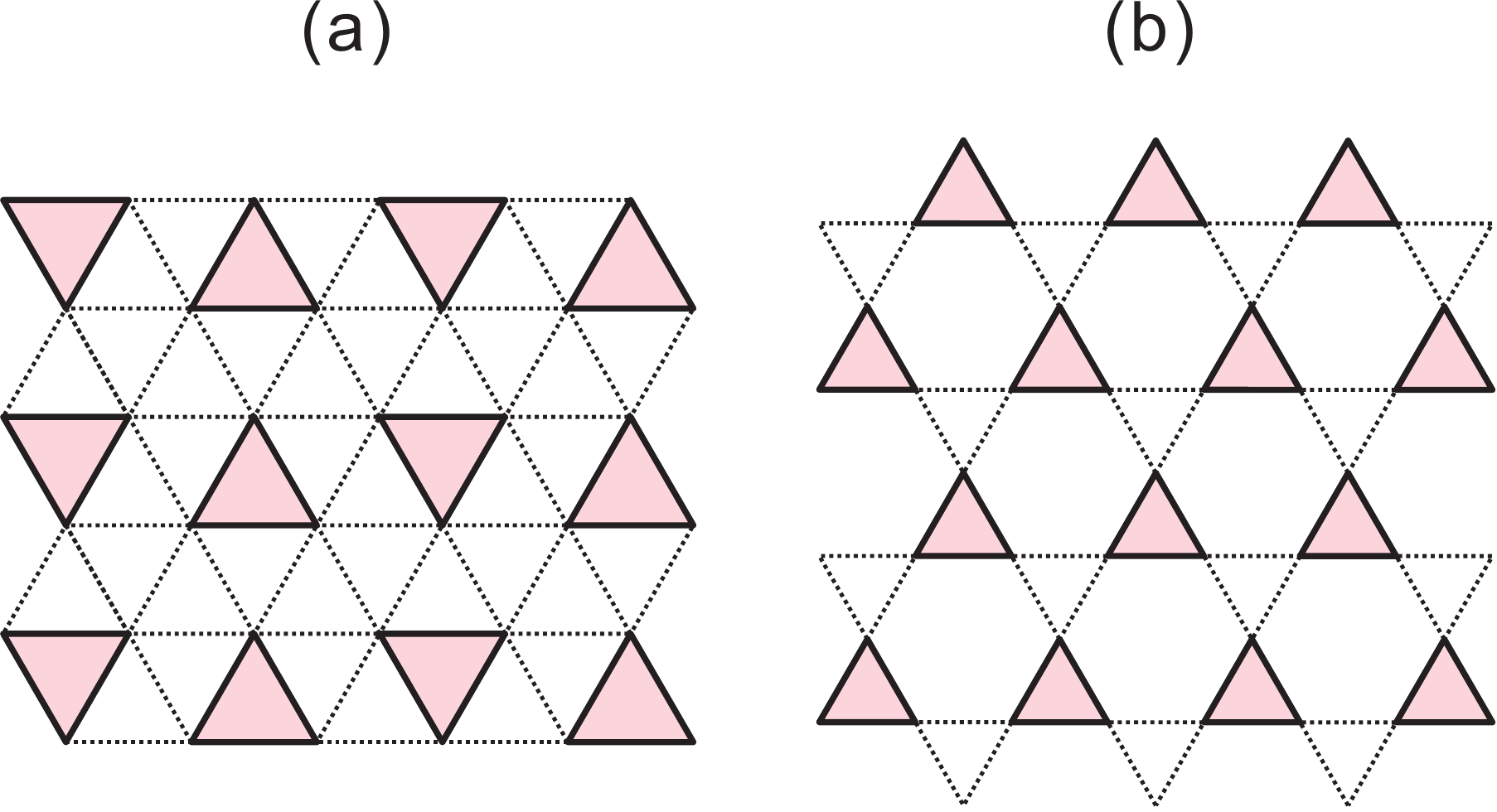}
\caption{
(a) Triangular and (b) kagome lattices as coupled triangles. Bold and dotted lines denote intra- and inter-triangle couplings, respectively.
}
\label{fig_2Dlattices}
\end{figure}

Many lattices can be described as coupled odd-dons. As examples, triangular and kagome lattices are illustrated in Fig.~\ref{fig_2Dlattices}(a) and (b), respectively. Note that the geometry of inter-triangle couplings in not unique in the triangular lattice. The other examples of coupled odd-gons are the shastry-Sutherland, pyrochlore lattices, and fullerenes, etc. As stated in the main text, a macroscopic ferromagnetism is expected to occur in the weak inter-odd-gon-coupling limit. It should be studied in future whether a spin-triplet superconductivity appears when the ferromagnetism is melted by increasing inter-odd-gon coupling.

\section{Finite-size scaling analysis of the bonding energy}

\begin{figure}[t]
\centering
\includegraphics[width=0.4\linewidth]{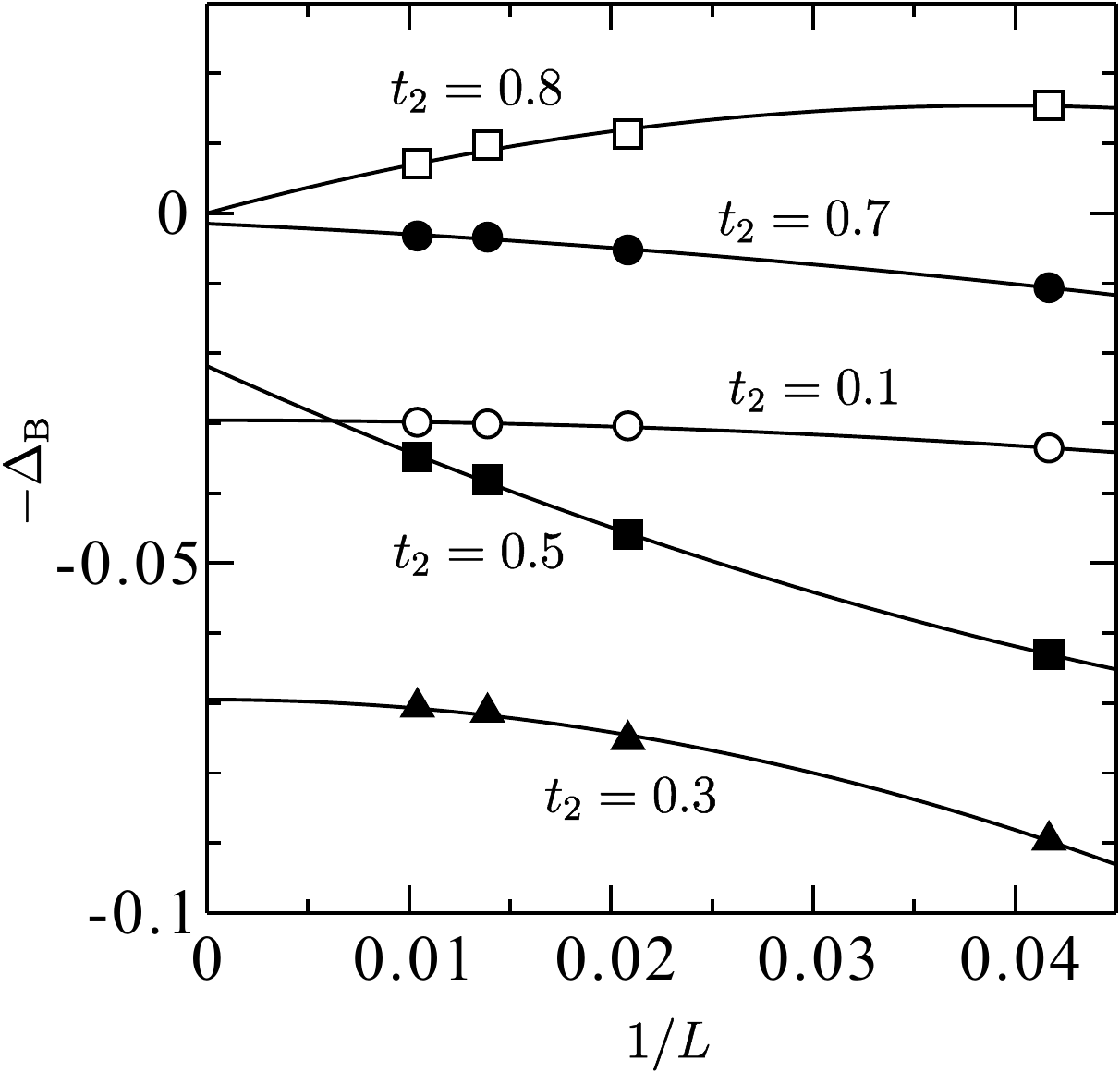}
\caption{
Finite-size scaling analyses of the binding energy at $n=7/12$.
}
\label{fig_scaling_DB}
\end{figure}

The binding energy of two fermions is defined as $\Delta_{\rm B}=\lim_{L\to\infty}\Delta_{\rm B}(L)$ with $\Delta_{\rm B}(L)=-[E_0(N_\uparrow\pm1,N_\downarrow\pm1;L)+E_0(N_\uparrow,N_\downarrow;L)-2E_0(N_\uparrow\pm1,N_\downarrow;L)]$, where $E_0(N_\uparrow,N_\downarrow;L)$ is the ground-state energy of the system with length $L$ containing $N_\uparrow$ spin-up and $N_\downarrow$ spin-down fermions. Since this quantity has a meaning only after being extrapolated to the thermodynamic limit $L\to\infty$, the finite-size scaling analysis is necessary. In Fig.~\ref{fig_scaling_DB}, some examples of the finite-size scaling analysis are shown.

\section{Superconducting pairing symmetry of A$_2$Cr$_3$As$_3$}

We investigated the pairing symmetry of our triplet superconductivity in the twisted triangular Hubbard tube. Symmetry of Cooper pairs give an invaluable information of our elucidating a superconducting mechanism. So far, the possibilities of {\it f}-wave and {\it p}$_z$-wave pairing symmetries have been suggested for the superconductivity in K$_2$Cr$_3$As$_3$. Our DMRG calculations were performed in real space. Therefore, it is more convenient to discuss the pairing symmetry in real space. Phenomenologically, the pairing Hamiltonian for one-dimensional system is written as
\begin{equation}
{\cal H}_{\rm pair}=\frac{1}{2N}\sum_{k,k^\prime,q,\sigma,\sigma^\prime}V_{\sigma,\sigma^\prime}(k,k^\prime)c_{k+q\sigma}^\dagger c_{-k\sigma^\prime}^\dagger c_{-k^\prime\sigma^\prime}c_{k^\prime+q\sigma},
\label{ham_pair}
\end{equation}
where $V_{\sigma,\sigma^\prime}(k,k^\prime)$ is the attractive interaction. For the $p$-wave superconductivity, the interaction is given by
\begin{equation}
V_{\sigma,\sigma^\prime}(k,k^\prime)=-V(\sqrt{2}\sin k)(\sqrt{2}\sin k^\prime),
\label{int_Vp}
\end{equation}
where $V$ is an averaged value of $V_{\sigma,\sigma^\prime}(k,k^\prime)$ around the Fermi level. Following the BCS theory, we apply a mean-field approximation to Eq.(\ref{ham_pair}). It leads to
\begin{equation}
{\cal H}_{\rm pair}^{\rm MF}\simeq\frac{1}{2N}\sum_{k,k^\prime,q,\sigma,\sigma^\prime}V_{\sigma,\sigma^\prime}(k,k^\prime) \langle c_{-k^\prime\sigma^\prime}c_{k^\prime+q\sigma}\rangle c_{k+q\sigma}^\dagger c_{-k\sigma^\prime}^\dagger + \langle c_{k+q\sigma}^\dagger c_{-k\sigma^\prime}^\dagger \rangle c_{-k^\prime\sigma^\prime}c_{k^\prime+q\sigma} - \langle c_{k+q\sigma}^\dagger c_{-k\sigma^\prime}^\dagger \rangle \langle c_{-k^\prime\sigma^\prime}c_{k^\prime+q\sigma}\rangle.
\label{ham_pair_MF}
\end{equation}
By usage, we define the superconducting order parameter (or gap function) as
\begin{equation}
\Delta_{\sigma,\sigma^\prime}(k,q)=\frac{1}{N}\sum_{k^\prime} V_{\sigma,\sigma^\prime}(k,k^\prime)\langle c_{-k^\prime\sigma^\prime}c_{k^\prime+q\sigma}\rangle
\label{SC_OP}
\end{equation}
Using Eq.(\ref{SC_OP}), the Hamiltonian (\ref{ham_pair_MF}) is rewritten as
\begin{equation}
{\cal H}_{\rm pair}^{\rm MF}=\frac{1}{2}\sum_{k,q,\sigma,\sigma^\prime}[\Delta_{\sigma,\sigma^\prime}(q)\sin k+h.c.]+\frac{N}{4V}\sum_{q\sigma\sigma^\prime}|\Delta_{\sigma\sigma^\prime}(q)|^2
\label{ham_pair_MF2}
\end{equation}
with
\begin{equation}
\Delta_{\sigma,\sigma^\prime}(q)=-\frac{2V}{N}\sum_{k^\prime}\sin k^\prime \langle c_{-k^\prime\sigma^\prime}c_{k^\prime+q\sigma}\rangle 
\label{SC_OP2}
\end{equation}
The Fourier transform of Eq.(\ref{ham_pair_MF2}) gives
\begin{equation}
{\cal H}_{\rm pair}^{\rm MF}=\sum_{r\sigma\sigma^\prime}[\Delta_{\sigma\sigma^\prime}(r)\frac{1}{2i}(c_{r\sigma}^\dagger c_{r+\hat{z}\sigma^\prime}-c_{r\sigma}^\dagger c_{r-\hat{z}\sigma^\prime})+h.c.]+\frac{1}{4V}\sum_{r\sigma\sigma^\prime}|\Delta_{\sigma\sigma^\prime}(r)|^2
\label{ham_pair_MF_rs}
\end{equation}
with the real-space order parameter
\begin{equation}
\Delta_{\sigma\sigma^\prime}(r)=-iV(\langle c_{r+\hat{z}\sigma^\prime}c_{r\sigma}\rangle-\langle c_{r-\hat{z}\sigma^\prime}c_{r\sigma}\rangle)
\label{SC_OP_rs}
\end{equation}

In order to determine the symmetry numerically, we calculated the pair field correlation function $D(i,j)=\langle \Delta_i \Delta_j^\dagger \rangle$ with $\Delta_i^\dagger=c_{i1\uparrow}^\dagger c_{i2\downarrow}^\dagger+c_{i1\downarrow}^\dagger c_{i2\uparrow}^\dagger$, where $c_{i\lambda\sigma}^\dagger$ creates a hole of spin $\sigma$ on a site $\lambda=1$ or $2$ of the bond indexed by $i$. The results are shown in Fig.~5 of the main text. From Eq.(\ref{SC_OP_rs}), $D(i,j)$ changes its sign alternately along the $z$-axis in the {\it p}$_z$ pairing symmetry.



\end{document}